\documentclass[a4paper,11pt]{article}
\pdfoutput=1
\usepackage{jcappub}

\usepackage{color}
\usepackage{amsfonts}
\usepackage{graphicx}
\usepackage{amssymb}
\usepackage{amsmath}

\usepackage{eufrak}
\usepackage{mathrsfs}

\renewcommand{\thesection}{\arabic{section}}

\def\laq{~\raise 0.4ex\hbox{$<$}\kern -0.8em\lower 0.62ex\hbox{$\sim$}~}
\def\gaq{~\raise 0.4ex\hbox{$>$}\kern -0.7em\lower 0.62ex\hbox{$\sim$}~}

\def\beq{\begin{equation}}
\def\eeq{\end{equation}}
\def\bea{\begin{eqnarray}}
\def\eea{\end{eqnarray}}

\def \pa {\partial}
\def \ra {\rightarrow}

\def \Mp {M_{\rm P}}

\def \Da {\Delta}
\def \da {\delta}
\def \b {\beta}
\def \a {\alpha}
\def \ap {\alpha^{\prime}}

\def \ga {\gamma}
\def \sg {\sigma}

\def \da {\delta}

\def \r {\rho}
\def \om {\omega}
\def \Om {\Omega}

\title{Constraints on the production \\of primordial magnetic seeds\\ in pre-big bang cosmology}

\author{M. Gasperini}

\affiliation{Dipartimento di Fisica, Universit\`a di Bari, \\ 
Via G. Amendola 173, 70126 Bari, Italy\\
and Istituto Nazionale di Fisica Nucleare, Sezione di Bari,\\ Via E. Orabona 4, 70125 Bari, Italy}

\emailAdd{gasperini@ba.infn.it}

\abstract{We study the amplification of the electromagnetic fluctuations, and the production of ``seeds" for the cosmic magnetic fields, in a class of string cosmology models whose scalar and tensor perturbations reproduce current observations and satisfy known phenomenological constraints. We find that the condition of efficient seeds production can be satisfied and compatible with all constraints only in a restricted region of parameter space, but we show that such a region has significant intersections with the portions of parameter space where the produced background of relic gravitational waves is strong enough to be detectable by aLIGO/Virgo and/or by eLISA.}

\keywords{primordial magnetic fields, 
primordial gravitational waves (theory), 
cosmological perturbation theory,
gravitational waves and CMBR polarization
 
\vskip13pt plus8pt minus11pt

\noindent{\bfseries\large\sffamily{Preprints:}} BA-TH/710-17
}

\begin{document}

\maketitle

%%%%%%%%%%%%%%%%%%%%%%%%%%%%%%%%%%%%%%%%%%%%%%%%%%%%%%%%%%%%%%%%%%%%%%%%%%%

\section{Introduction}
\label{Sec1}
\setcounter{equation}{0}

In a previous paper \cite{1} we have discussed a class of string cosmology models able to produce -- via the axion/curvaton mechanism -- a spectrum of scalar metric perturbations in agreement with all results obtained from the most recent CMB observations (see e.g \cite{2,3,4}). We have shown \cite{1}, also, that the stochastic background of pre-big bang gravitational waves (GW) predicted by these models \cite{5,6,7} may be compatible with all present phenomenological constraints and, in an appropriate region of parameter space, may be detectable by the interferometers aLIGO/Virgo and eLISA, operating at their final design sensitivity \cite{8,9}. 

In this paper we address the question of whether, in the same class of models, the inflationary amplification of the electromagnetic (e.m.) vacuum fluctuations may be efficient enough to provide the ``seeds" needed to generate -- via the galactic dynamo mechanism -- the magnetic fields currently observed on large scales (see e.g. \cite{10,11}). Such a possibility has already been discussed in the context of pre-big bang cosmology \cite{12,13,14,15}, without taking into account, however, the full set of constraints imposed by the associated production of a viable spectrum of scalar and tensor perturbations. 

By considering, in particular, the ``minimal" version of the so-called pre-big bang scenario \cite{16,17} it will be shown that, in addition to the difficulties pointed out in \cite{17a}, an efficient production of e.m. seeds is incompatible with the  phenomenological  limits following from the CMB radiation and the stochastic GW background, at least if we restrict ourselves to the peculiar photon-dilaton coupling predicted by the heterotic string model in the low-energy and weak coupling limit (i.e. the coupling adopted in previous papers \cite{12,13,14,15}). 

However, if we allow for a more general parametrization of the coupling of the photon to the dilaton (and to the internal moduli fields), then we can find regions of parameter space where all phenomenological constraints are satisfied, where the produced e.m. fluctuations are large enough to seed the galactic dynamo, and, simultaneously, the relic GW background is possibly detectable by the existing interferometric antennas. We may conclude that future, cross-correlated results from CMB observations and GW detectors can provide information, in principle, on the possible origin of the cosmic magnetic fields and also, indirectly, on the effective couplings of the string theory sector reproducing the standard e.m. interactions.

The paper is organized as follows. In section \ref{Sec2} we present our class of models and our assumptions on the basic parameters controlling the evolution of the cosmological background. In section \ref{Sec3} we recall the full set of phenomenological constraints (derived in \cite{1}) needed to obtain a viable spectrum of scalar and tensor perturbations, and we define the allowed region in the corresponding parameter space. In section \ref{Sec4} we discuss the amplification of the e.m. fluctuations and the efficient production of seeds for the cosmic magnetic fields, for different values of the parameter controlling their coupling to the dilaton/moduli background. After imposing all relevant constraints, we compare the allowed region of parameter space corresponding to seed production with the allowed region corresponding to a detectable GW background. In section \ref{Sec5} we present our final remarks and conclusions. 

%%%%%%%%%%%%%%%%%%%%%%%%%%%%%%%%%%%%%%%%%%%%%%%%%%%%%%%%%%%%%%%%%%%%

%%%%%%%%%%%%%%%%%%%%%%%%%%%%%%%%%%%%%%%%%%%%%
\section{The model of background evolution and its basic parameters}
\label{Sec2}
\setcounter{equation}{0}
%%%%%%%%%%%%%%%%%%%%%%%%%%%%%%%%%%%%%%%%%%%%%%%%

The cosmological scenario of this paper is the same as the one considered in \cite{1}: a higher-dimensional string background evolves from the string perturbative vacuum through a phase of dilaton-driven, ``pre-big bang" inflation, reaches the high-curvature and strong coupling regime, and eventually decays towards a state of standard, ``post-big bang", radiation-dominated expansion \cite{16,17}. The transition from the pre- to the (dual) post-big bang regime is associated with a regular bounce of the space-time curvature \cite{18,19}, is expected to trigger the stabilization of the dilaton and of the internal moduli \cite{20}, and is possibly followed by a dust-like  epoch of decelerated evolution, dominated by the non-relativistic oscillations of the Kalb-Ramond axion background \cite{21,22}.

For the computation of the associated spectrum of cosmological perturbations we need to specify the kinematical details (as well as the duration, and the relative curvature scales) of the various phases of background evolution. Let us start with the pre-big bang regime.

We shall work with a simple model of background geometry with three isotropically expanding dimensions and six (internal) shrinking dimensions, not necessarily isotropic, described in the so-called string frame by the spatially flat metric
\beq
ds^2= dt^2 - a^2(t) |d\vec x|^2 - \sum_i b_i^2(t) dy_i^2, ~~~~~~~~~~~~~ i=1, \dots, 6.
\label{21}
\eeq
This geometry is sourced by a dynamical dilaton field $\phi(t)$, and satisfies the cosmological equations following from the string effective action (see e.g. \cite{17,23}). The axion and the e.m background fields are trivial, $\sg=0$, $A_\mu=0$ (but their quantum fluctuations are non-vanishing). The conformal time coordinate $\eta$ is related to the cosmic time $t$, as usual, by $dt=a d\eta$). 

We will consider, in particular, a minimal model of pre-big bang (i.e. pre-bouncing) evolution, with two phases: an initial low energy and weak coupling phase, followed by a more ``stringy", high-curvature phase whose properties crucially depend on the inclusion of the so-called $\ap$ corrections into the string effective action \cite{24}.

In the initial low-curvature phase, ranging in conformal time from $-\infty$ to a transition time scale $\eta=-\eta_s<0$, the background (\ref{21}) can be appropriately described by a vacuum solution of the tree-level string cosmology equations  \cite{25}, with a Bianchi-I type metric given by
\beq
a(\eta) \sim (-\eta)^{\b_0\over 1- \b_0}, ~~~~~~~~ b_i(\eta) \sim  (-\eta)^{\b_i\over 1- \b_0}, ~~~~~~~~~~ \eta<-\eta_s,
\label{22}
\eeq
and with 
\beq
\phi(\eta) \sim { \sum_i \b_i +3 \b_0 -1\over 1-\b_0} \ln (-\eta),
~~~~~~~~~~~~~~~~~~~~~~~ \eta<-\eta_s,
\label{23}
\eeq
where $\b_0$, $\b_i$ are constant parameters satisfying the Kasner-like condition
\beq
3 \b_0^2+ \sum_i \b_i^2=1.
\label{24}
\eeq
In the subsequent, high-curvature, string phase, ranging from $\eta=-\eta_s$ to the final strong-coupling scale $\eta=-\eta_1$ (with $\eta_s>\eta_1$), our background geometry (\ref{21}) is described instead by a fixed-point solution of the string cosmology equations with the appropriate $\ap$ corrections \cite{24}. For such solutions, which represent late-time attractors of the preceding low-energy evolution, the space-time curvature stays frozen at a constant scale $H_1$ related to the value of the fundamental string mass parameter, and always smaller than Planckian, $H_1< \Mp= (8 \pi G)^{-1/2}$. The effective four-dimensional string coupling, 
\beq
g= \left(\prod_i  b_i\right)^{-1/2} e^{\phi/2},
\label{25}
\eeq
on the contrary, keeps monotonically growing, and its evolution, in conformal time, can be parametrized by a simple power-law behavior. Hence, for such solutions \cite{24},
\beq
a(\eta) \sim (-\eta)^{-1}, ~~~~~~ g(\eta) \sim (-\eta)^{-\b},  ~~~~~~~~~~~~~~
-\eta_s < \eta <-\eta_1,
\label{26}
\eeq
where the parameter $\b$ is not completely free, but it is constrained to be in the range
\beq 
0 \leq \b < 3.
\label{27}
\eeq
The lower limit on $\b$ is due to our assumption of growing string coupling (needed to implement a successful bouncing transition), while the upper limit is to be imposed to avoid background instabilities \cite{26}.

Summarizing we can say that, for the given model of pre-big bang evolution and for the computation of the perturbation spectra of interest for this paper, the relevant parameters (controlling amplitude, slope, and frequency positions of the different branches of the various spectra) are four: $\b_0$, $\b$, $\eta_s$ and $\eta_1$ (see also \cite{1} and the following sections). The last two parameters can be conveniently replaced by two equivalent, more physical, quantities: the Hubble scale $H_1$,
associated  with the curvature of the string phase, and the redshift parameter $z_s=a_1/a_s= \eta_s/\eta_1$, describing the expansion of the $3$-dimensional space during that phase. Finally, the parameter $\b$ can also be expressed in terms of the overall growth of the four-dimensional string coupling during the string phase, using the relation 
\beq
g_s/g_1=( \eta_s/\eta_1)^{-\b} \equiv z_s^{-\b}.
\label{28}
\eeq

Hence, we end up with the following set of relevant pre-big bang parameters: $\b_0$, $H_1$, $z_s$ and $g_s/g_1$. As we shall see in the following section, two of these parameters (in particular, $\b_0$ and $H_1$) can be fixed and/or expressed in terms of the other variables, using the stringent constraints imposed by CMB data. We will thus reduce to a class of models with a two-dimensional parameters space, spanned by the variables $z_s$ and $g_s/g_1$.

In the subsequent post-big bang regime, starting after the bounce (which, for the applications of this paper, can be assumed to be almost instantaneously localized at the transition epoch $\eta=-\eta_1$), the curvature of our background is monotonically decreasing, the dilaton and the extra spatial dimensions are frozen, and the Universe is filled with a hot radiation gas produced by the transition.
In addition, the Kalb-Ramond axion emerges from the bounce with a mass $m$ and a non-trivial background value $\sg_i \not=0$, displaced from the minimum of the (non-perturbative, periodic) axion potential. 
Excluding the exotic ``trans-Planckian" possibility $\sg_i > \Mp$, (which would include a phase of axion-dominated, post-big bang, slow-roll inflation \cite{22}), the effective axion potential can be assumed to take the form of a quadratic mass term. In such a case the axion starts oscillating with proper frequency $m$ as soon as  the curvature drops below the scale $H_m \sim m$. It then behaves like a dust fluid, its energy grows with respect to the radiation energy density, and eventually dominates the cosmic expansion at the curvature scale $H_\sg \sim m (\sg_i/\Mp)^4$  \cite{21,22}. 

We will assume that the Kalb-Ramond axion is minimally coupled to gravity and that it decays with gravitational coupling strength, disappearing from the cosmological scene at a decay scale $H_d \sim m^3/\Mp^2$ \cite{21,22}. To avoid disturbing the standard nucleosynthesis scenario we must impose on the decay to occur prior to nucleosynthesis, i.e. at a scale $H_d >H_N \sim (1\, {\rm MeV})^2/\Mp$. It is important to mention, finally, that the possible coupling of the axion to other gauge fields, if not subdominant, might lead to redefine the assumed decay scale $m^3/\Mp^2$ and change, in principle, the epoch (and the duration) of the axion-dominated evolution. A study of how this possibility might affect the class of backgrounds considered here is postponed to a future paper. 

As discussed in \cite{1}, the above model of post-big bang evolution is consistent, and can successfully convert the axion fluctuations amplified by pre-big bang inflation into a viable spectrum of adiabatic scalar metric perturbations, provided the following hierarchy of curvature scales is satisfied: 
\beq
H_1\gaq H_m \gaq H_\sg \gaq H_d > H_N, 
\label{29}
\eeq
where $H_1<\Mp$.
The background evolution is of the matter-dominated type ($a \sim \eta^2$) in the axion phase ranging from $H_\sg$ down to $H_d$, and of the radiation-dominated type ($ a\sim \eta$) in all other phases. The regime of post-big bang evolution thus provide us, in principle, with two more parameters to add to those determining the shape and the amplitude of the metric perturbation spectra: the axion mass $m$ and the initial axion amplitude $\sg_i$. They control the position in time and the duration of the different axion phases of the cosmic dynamics.

In practice, however, the particular values of $m$ and $\sg_i$ have a much smaller impact on the discussion (and on the final results) of this paper than the values of the  pre-big bang parameters $\b_0$, $H_1$, $z_s$, $g_s/g_1$ (see also \cite{1}). We will thus restrict the present analysis to the same class of models as the one considered in \cite{1}, characterized by the following (natural) 
values of the axion parameters: 
\beq
m=H_1, ~~~~~~~~~~~~~~~~~~ \sg_i=\Mp.
\label{210}
\eeq
In that case $H_\sg=H_m=H_1$, the decay scale is given by $H_d= (H_1^3/\Mp^2)$, and the consistency condition (\ref{29}) simply reduces to the following constraint on the string curvature scale $H_1$:
\beq
\Mp>H_1\gaq 10^{-14} \Mp.
\label{211}
\eeq

Let us finally complete the definition of the class of models considered in this paper with two further (convenient) conditions. 

First of all we shall impose that the extension of the high-curvature string phase towards the past is limited in time, in such a way that all (low-frequency) perturbation modes affecting the (large-scale) distances relevant to the observed CMB anisotropy leave the horizon during the initial, low-energy, pre-big bang regime. More precisely, let us assume that
$
\om_\ast <\om_s
$. 
Here $\om_\ast$ is the proper frequency corresponding to the pivot scale $k_\ast = 0.05$ Mpc$^{-1}$ to which CMB measurements are typically referred (see e.g. \cite{4}), while $\om_s= H_s a_s/a$ is the proper frequency of a mode leaving the horizon just at the epoch $\eta=-\eta_s$, marking the end of the low energy dilaton-driven phase and the beginning of the high-curvature string phase. (Throughout this paper we will work with the proper frequency $\om$, related to the Fourier parameter $k$ by $\om= k/a$. Also, we will use the shortened notation $H_s\equiv H(\eta_s)$, $a_s \equiv a(\eta_s)$, while $a \equiv a(\eta)$). By computing the ratio $\om_\ast/\om_s$ for our model of background evolution we thus obtain the condition 
\beq
z_s\left(H_1/\Mp\right)^{-5/6} \laq 10^{27}.
\label{212}
\eeq
(see \cite{1} for a detailed computation).

Second, we shall assume that the post-big bang regime of axion-dominated evolution may possibly affect (after they have re-entered the horizon) only those high-frequency modes which have left the horizon during the string phase. In other words, we shall assume that $\om_s<\om_d$, where $\om_d= H_d a_d/a$ is the proper frequency of a mode re-entering the horizon at the epoch of axion decay, when $H=H_d$. By expressing the ratio $\om_s/\om_d$ in terms of our parameters (see \cite{1} for a detailed computation) we 
thus obtain the condition
\beq
z_s\left(H_1/\Mp\right)^{2/3} \gaq 1,
\label{213}
\eeq
which complete the set of restrictions needed to specify the class of models of our interest.

It is important to stress that relaxing the conditions (\ref{210}), (\ref{212}), (\ref{213}) on the given set of pre- and post-big bang parameters would complicate the analysis of the perturbation spectra but, as discussed in \cite{1}, would not enhance in any significant way the allowed region of parameter space compatible with GW detection (at least, for the limiting sensitivities and the frequency bands accessible to currently operating interferometers). This means, in other words, that the one we are considering here is the {\em ``simplest"} among the {\em ``most promising"} classes of models able to extract significant information on string cosmology/pre-big bang dynamics from the forthcoming data of the gravitational antennas.

%%%%%%%%%%%%%%%%%%%%%%%%%%%%%%%%%%%%%%%%%%%%%%%%%%%%%%%%%%%%%%%%%%%%%%%%%%%

%%%%%%%%%%%%%%%%%%%%%%%%%%%%%%%%%%%%%%%%%
\section{Phenomenological constraints from scalar and tensor perturbations}
\label{Sec3}
\setcounter{equation}{0}
%%%%%%%%%%%%%%%%%%%%%%%%%%
In this section we will introduce the set of constraints to be imposed on the model of section 2 in order to obtain a viable spectrum of metric perturbations, and we will determine the allowed region of the corresponding parameter space.

Let us first recall that, in general, the amplification of a cosmological perturbation $\da$ is controlled by the so-called ``pump field" $\xi(\eta)$, associated to the canonical variable $u=\xi \da$ whose Fourier components $u_k$ satisfy the standard mode equation \cite{27}
\beq u_k'' +\left(k^2 -\xi''/\xi\right) u_k=0
\label{31}
\eeq
(a prime denotes differentiation with resect to the conformal time $\eta$). Let us consider an accelerated (inflationary) phase in which the pump field evolves in time as $\xi(\eta) \sim (-\eta)^\a$ , $\eta<0$, $\eta \ra 0_-$. Once the above equation is satisfied, the primordial (super-horizon) distribution of the perturbation modes $\da_k= u_k/\xi$, normalized to a vacuum fluctuation spectrum, and leaving the horizon during the given phase of accelerated evolution, is described by the so-called spectral amplitude $\Da^2(k)$, and is asymptotically controlled by the power $\a$ of the pump field as \cite{27}
\beq
\Da^2(k) \equiv {k^3 \over 2 \pi^2} \left|\da_k\right|^2 \sim k^{3- |2 \a-1|}.
\label{32}
\eeq

It is important to stress that the canonical equation (\ref{31}) and the pump field $\xi$, for each type of perturbation, must be explicitly determined by perturbing and diagonalizing the effective action describing the associated dynamics. In our scenario of pre-big bang inflation there are two phases, and the dynamics of the high-curvature string phase is described by an action which includes $\ap$ corrections and which is different, in general, from the low-energy string effective action. However, the particular solution representing  the string phase corresponds to a (fixed-point) background configuration characterized by constant values of the Hubble parameters and linear evolution (in cosmic time) of the dilaton field \cite{24}:  in that case the standard definitions of the pump fields (as well as the canonical low-energy form of the perturbation equation) may be safely extrapolated to the high-curvature regime. This result has been explicitly checked for the case of tensor metric perturbations  \cite{28}, and it is  expected to hold for all types of perturbations, at least in the linear approximation. 

By applying the standard formalism of linear perturbation theory let us then consider the pre-big bang amplification of the vacuum fluctuations of the Kalb-Ramond axion, and the related production (through the curvaton mechanism \cite{29,30,31}) of a super-horizon (post-big bang) spectrum of adiabatic scalar metric perturbations. Following the results of previous computations \cite{21,22}, and recalling that in our case, as specified by the constraints (\ref{212}), the frequency modes relevant to CMB observations are those leaving the horizon during the low-energy phase, we find that the primordial spectral amplitude of curvature perturbations is given by \cite{1}
\beq
\Da_{\cal R}^2(\om)=
 {f^2(\sg_i)\over 2 \pi^2} \left(H_1\over \Mp\right)^2
 \left(\om_s\over \om_1\right)^{-2\b}\left(\om\over \om_s\right)^{3 -2\left|3 \b_0-1\over 1-\b_0\right|}, 
~~~~~~~~~~~
 \om<\om_s.
 \label{33}
 \eeq
As usual, $\om_1 =H_1 a_1/a$ and $\om_s =H_s a_s/a$ are the proper frequencies of  modes crossing the horizon, respectively, at the end ($\eta=-\eta_1$) and at the beginning ($\eta=-\eta_s$) of the string phase (we may note that $H_s=H_1$, hence the ratio $\om_1/\om_s=a_1/a_s=z_s$ defines the redshift parameter related to the time extension of this phase). Finally, $f(\sg_i)$ is the transfer function connecting axion fluctuations to scalar metric perturbations: its explicit form in terms of $\sg_i$ has been numerically computed in \cite{22}, and for the particular case $\sg_i=\Mp$ considered here we have $f^2(\Mp) \simeq 0.137$.

By using the observed value of the scalar spectral index \cite{2,4}, $n_s\simeq 0.968$, we can now fix the parameter $\b_0$ by imposing the condition
\beq
3 -2\left|3 \b_0-1\over 1-\b_0\right| \equiv n_s-1 .
\label{34}
\eeq
To resolve this equation for $\b_0$ we have to take into account that, for the consistency of our cosmological scenario, the value of $\b_0$ must be compatible with the growth in time of the effective four-dimensional string coupling (\ref{25}). To this purpose we insert the background solution (\ref{22})--(\ref{24}) into Eq. (\ref{25}),  and we find that the evolution of $g$ during the low energy, dilaton-driven phase is given by
\beq g(\eta) \sim (-\eta)^{{1\over 2} {3 \b_0-1\over 1-\b_0}},  ~~~~~~~~~~~~~~ \eta< -\eta_s.
\label{35}
\eeq
The coupling is growing, for $\eta \ra 0_-$, provided that $(3 \b_0-1)/( 1-\b_0))<0$. Using this condition we immediately obtain, from Eq. (\ref{34}),
\beq
\b_0 = {n_s-2\over n_s+2}, ~~~~~~~~~~~~~~~~~ n_s \simeq 0.968.
\label{36}
\eeq

An additional, important constraint on the parameters of our scenario follows from the normalization of the spectral amplitude (\ref{33}) at the pivot scale $\om_\ast$, which imposes the experimental condition \cite{2,4} $\Da_{\cal R}^2 (\om_\ast)\simeq 3 \times 10^{-10}$. Using Eq. (\ref{34}) for the spectral index, we can express the above normalization condition as \cite{1}
\beq
\left(H_1\over \Mp\right)^{17- 5 n_s\over 6} \simeq z_s^{1-n_s-2\b}  \times {6 \pi^2 \over f^2(\Mp)} \times 10^{27 n_s- 37}, ~~~~~~~~~~
f^2(\Mp) \simeq 0.137.
\label{37}
\eeq
As anticipated in Sect. 2, we will use this condition to eliminate everywhere $H_1$ in terms of the other two parameters $\b$ and $z_s$.

Let us now consider the phenomenological constraints following from the amplification of tensor perturbations and the production of a stochastic background of relic GW radiation. The today value of its spectral energy density, $\Om_{GW}(\om, t_0)$, in critical units, can be written as \cite{1}
\bea
\Om_{GW}(\om, t_0) &=& 
\Om_r (t_0) \left(H_1\over \Mp\right)^2 
\left(\om\over \om_1\right)^{1-|3-2\b|},
~~~~~~~~~~~~~~~~~~~~~~~~~~~~~~~~ \om_d<\om<\om_1,
\nonumber \\ &=&
\Om_r (t_0) \left(H_1\over \Mp\right)^2 
\left(\om_d\over \om_1\right)^{1-|3-2\b|}
\left(\om\over \om_d\right)^{3-|3-2\b|},
~~~~~~~~~~~~~~ \om_s<\om<\om_d,
\nonumber \\ &=&
\Om_r (t_0) \left(H_1\over \Mp\right)^2 
\left(\om_d\over \om_1\right)^{1-|3-2\b|}
 \left(\om_s\over \om_d\right)^{3-|3-2\b|} \left(\om\over \om_s\right)^3,
~~~~~ \om<\om_s.
\label{38}
\eea
Here $\Om_r(t_0)$ is the present value of the total fraction of critical energy density in the form of cosmic radiation (with dominant photon and neutrino components, see \cite{4} for its precise numerical value). As before, modes with $\om<\om_s$ leave the horizon during the low-energy, dilaton-driven phase, while modes with $\om >\om_s$ leave the horizon during the string phase. Finally, high-frequency modes in the band $\om_d<\om<\om_1$  re-enter the horizon during the axion-dominated (post-big bang) phase, and the slope of their spectrum is tilted towards the red (with respect to the super-horizon spectrum) by an additional $\om^{-2}$ factor \cite{1}.

A first phenomenological constraint on this spectrum comes from nucleosynthesis \cite{32}, not to spoil the accurate predictions on the abundance of light elements.  It can be expressed as  a crude upper limit on the peak intensity of the spectrum,
\beq
\Om_{GW}(\om_{\rm peak}) \laq 10^{-1}  \Om_r(t_0),
\label{39}
\eeq
to be imposed at the peak frequency $\om_{\rm peak}$ (which may correspond  either to $\om_d$ or to $\om_1$, depending on the value of the parameter $\b$).
Notice that the numerical value of the above limit practically coincides with the experimental upper bound recently placed by the LIGO and Virgo data \cite{33} on the amplitude of $\Om_{GW}$, in the frequency band of 41--169 Hz. 

A second, well known condition comes from the observations of millisecond pulsars \cite{34}: in particular, from the absence of any detectable distortion of pulsar timing due to the presence of a stochastic GW background, at a frequency scale $\om_p$ of the order of $10^{-8}$ Hz. This gives the bound
\beq
\Om_{GW}(\om_{p}) \laq 10^{-8} .
\label{310}
\eeq

A third important condition, for the class of models we are considering, comes from the request that the scalar metric perturbations directly amplified by the phase of pre-big bang inflation (and characterized by a primordial spectrum $\Da_\psi^2$ closely related to the graviton spectrum (\ref{38}), with a ``wrong" blue slope),  be negligible with respect to the spectrum of adiabatic scalar perturbations $\Da_{\cal R}^2$ obtained from the axion (and given by Eq. (\ref{33})). The condition $\Da_\psi^2(\om)< \Da_{\cal R}^2(\om)$ has to be satisfied in particular at the pivot scale $\om_\ast$, but, more generally, it has to be imposed at the maximum frequency scale $\om_M$ interested by the multipole expansion of the CMB anisotropy and constrained by present observations, namely at $\om =\om_M \simeq 6 \om_\ast$. This gives the condition \cite{1}:
\beq
\left(H_1\over \Mp\right)^2 
\left(\om_d\over \om_1\right)^{1-|3-2\b|}
 \left(\om_s\over \om_d\right)^{3-|3-2\b|}
\laq 6 \pi^2 \times 10^{-10} \left(\om_M\over \om_s\right)^{n_s-4}.
\label{311}
\eeq
We have checked that this condition is slightly stronger than a similar constraint imposed on the ratio $r$ of the tensor-to-scalar spectral amplitude, and obtained by the recent measurements of the CMB polarization \cite{3}, which imply, at the pivot scale, $r_{0.05} \equiv \Da_h^2(\om_\ast)/\Da_R^2(\om_\ast)<0.07$.

Let us recall, finally, that a cosmic stochastic background of relic GW radiation is accessible to the maximal  sensitivity of the aLIGO-AdVirgo detector network (expected to be reached in the 2020 by the so-called O5 observing run \cite{8}), provided that
\beq
\Om_{GW}(\om_L) \gaq 10^{-9} ,
\label{312}
\eeq
where $\om_L \simeq 10^2$ Hz. Also, the GW background is expected to be detectable by eLISA (in the so-called C1 configuration \cite{9}) provided that 
\beq
\Om_{GW}(\om_{eL}) \gaq 10^{-13} ,
\label{313}
\eeq
where $\om_L \simeq 10^{-2}$ Hz. 

\begin{figure}[t]
\centering
\includegraphics[width=9cm]{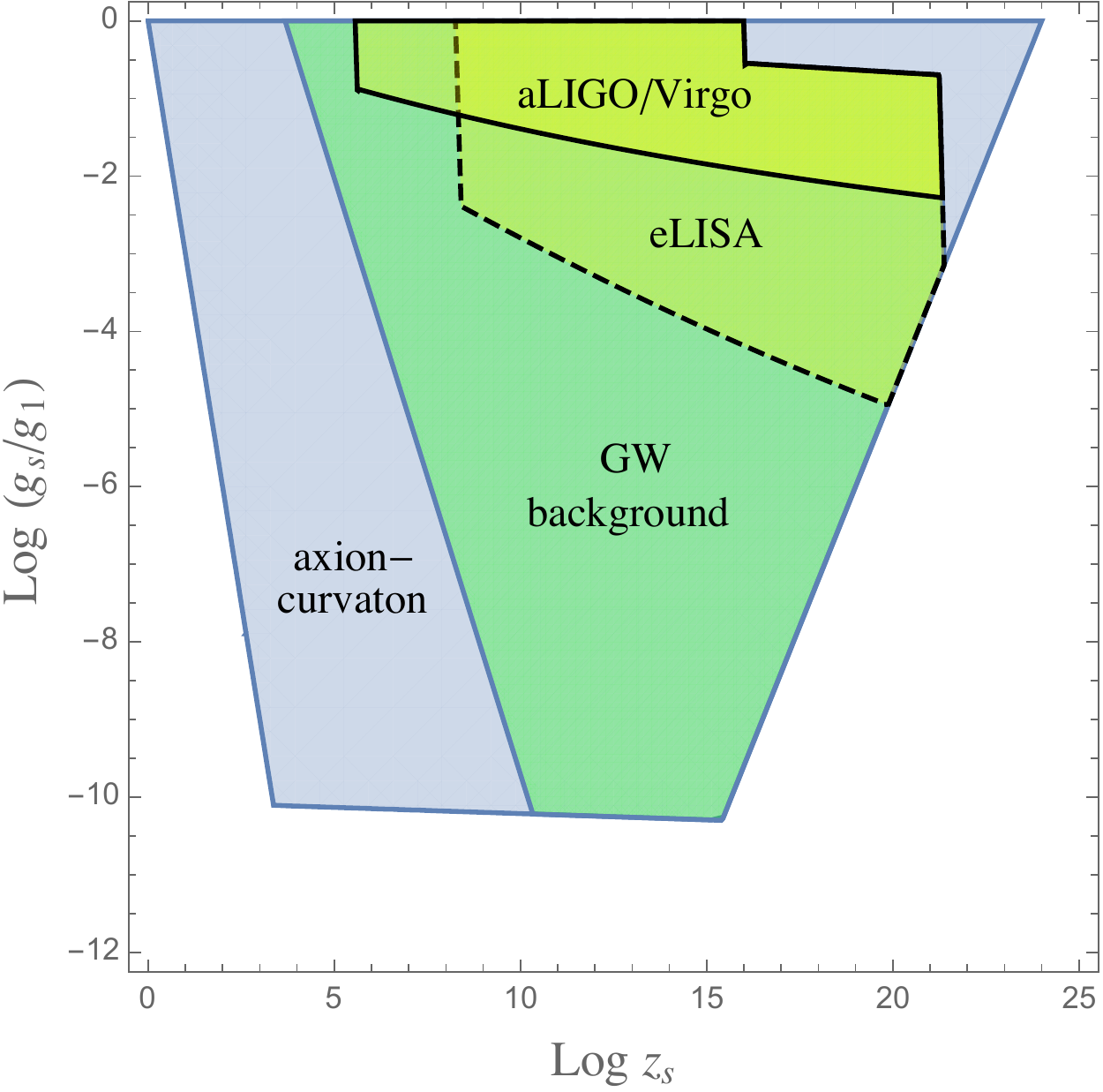}
\caption{The allowed region of parameter space for the production of a stochastic GW background which satisfies all phenomenological bounds (green area), and which is contained inside the larger, light-blu trapezoidal area compatible with the production of a viable spectrum of scalar perturbations via the axion/curvaton mechanism. Also shown in the figure are the smaller allowed regions where the produced GW background is directly detectable by aLIGO/Virgo (yellow area with thick border) and by eLISA (yellow area with dashed border).} 
\label{f1}
\end{figure}

We are now in the position of combining  all conditions presented here and in the previous section, in order to define  the corresponding allowed region of parameter space. After expressing Eqs. (\ref{39})--(\ref{313}) in terms of the parameters of our background (like the other constraints), we fix $\b_0$ according to Eq. (\ref{36}) and eliminate $H_1$ everywhere using Eq. (\ref{37}). The given set of constraints then defines an allowed area in the plane spanned by the two independent variables $z_s$ and $\b$ (controlling, respectively, the duration of the string phase and the associated growth of the string coupling). For graphical reasons it is convenient to replace $\b$ with the ratio $g_s/g_1$ (according to Eq. (\ref{28})),  and  switch to logarithmic variables defined by
\beq x= \log z_s, ~~~~~~~~~~~~~~~~~~
y= \log (g_s/g_1) =- \b x. 
\label{314}
\eeq
By expressing all constraints in terms of $x$ and $y$ (see also \cite{1}) we then obtain the results illustrated in Fig. \ref{f1}.

The green area of the $(x,y)$ plane represents the allowed region of parameter space for a viable GW background which satisfies the constraints imposed by Eqs. (\ref{213}), (\ref{39})--(\ref{311}) and which, in addition, is compatible with the production of a viable spectrum of scalar metric perturbations (namely, which is inside the region represented by the larger, light-blu trapezoidal area determined by the constraints (\ref{27}), (\ref{211}), (\ref{212}),  (\ref{37})). In the figure we have also represented, with the yellow regions included within the green area, the regions of parameter space compatible with the production of a GW background which satisfies all constraints and is directly detectable by aLIGO/Virgo (the yellow area with the thick border, satisfying the condition (\ref{312})), or by eLISA (the yellow area with the dashed border, satisfying the condition (\ref{313})).

In the following section we will discuss the possibility that the allowed region of parameter space corresponding to the efficient production of a viable spectrum of seeds for the cosmic magnetic fields may have (or not) some intersection with the green and yellow regions of Fig. 1.

%%%%%%%%%%%%%%%%%%%%%%%%%%%%%%%%%%%%%%%%%%%%%%%%%%%%%%%%%%%%%%%%%%%%%%%%%%%
%%%%%%%%%%%%%%%%%%%%%%%%%%%%%%%%%%%%%%%%%%%%%%%%%%%%%%%%%%%%%%%%%%%%%%%%%%%

%%%%%%%%%%%%%%%%%%%%%%%%%%%%%%%%%%%%%%%%%%%%%%%
\section{The amplifications of electromagnetic fluctuations}
\label{Sec4}
\setcounter{equation}{0}
%%%%%%%%%%%%%%%%%%%%%%%%%%%%%%%%%%%%%%%%%%%%%%%

Let us now discuss the amplification of the vacuum fluctuations of the e.m. field, considering the same class of cosmological models as before. The unperturbed e.m. background is vanishing, and the dynamics of its four-dimensional fluctuations $\da A_\mu$ is described by the   (dimensionally reduced) effective action
\beq
S_\ga= -{1\over 4} \int d^4x \sqrt{-g} \,\xi_\ga^2(b_i, \phi) F_{\mu\nu}F^{\mu\nu},
\label{41}
\eeq
where $F_{\mu\nu}= \pa_\mu \da A_\nu- \pa_\nu \da A_\mu$, and where the time-dependent function $\xi_\ga$ depends on the dilaton $\phi$ and on the internal moduli $b_i$. 
Working with free e.m. radiation in a spatially flat geometry we can conveniently adopt for the four-dimensional metric the conformal time gauge, $g_{\mu\nu} = a^2 \eta_{\mu\nu}$, and impose on the fluctuations the radiation gauge, $\da A_0=0$, $\pa^i \da A_i =0$. The action for the fluctuations $\da A_i$ can then be recast in a diagonal form by introducing, for each of the two independent polarization modes, the canonical variables $u_i = \xi_\ga \da A_i$. Their Fourier components satisfy Eq. (\ref{31}) with $\xi= \xi_\ga$.

The explicit form of the pump field $\xi_\ga$ depends on the type of superstring we are considering, and on the particular sector (Ramond or Neveau-Schwarz) containing the field to be identified with the gauge potential of the e.m. $U(1)$ symmetry. For instance, for the heterotic, type I and type IIA superstrings the coupling to the internal moduli arises, at low energy, from the coupling of $F_{\mu\nu}$ to the volume of the extra-dimensional manifold ($V_6= \prod_i b_i$), and for the heterotic model, in particular, one finds $\xi_\ga= g^{-1}$, where $g$ exactly corresponds to the effective string coupling of Eq. (\ref{25}) (see e.g. \cite{12,13,14,15}). 

In this paper we will adopt a phenomenological approach, assuming that the effective pump field can be parametrized as
\beq
\xi_\ga= \left[\left(\prod_i  b_i\right)^{1/2} e^{-\phi/2}\right]^\ga,
\label{42}
\eeq
where $\ga$ is a constant. The value $\ga=1$ reproduces the case of the heterotic model in the low-energy limit. The case $\ga \not=1$ parametrizes a different effective coupling of $F_{\mu\nu}$ to the dilaton and/or to the internal moduli, possibly reproducing the results of different models of superstrings.

Using the model of background illustrated in section \ref{Sec2} we immediately find that the pump field (\ref{42}) evolves in time as 
\beq
\xi_\ga(\eta) \sim \left(-\eta \right)^{{\ga\over 2}{1-3 \b_0\over 1- \b_0}}, ~~~~~~~~~~~~~~~~~~~
\eta <-\eta_s,
\label{43}
\eeq
during the initial low energy phase described by the solution (\ref{22}), (\ref{23}), and as
\beq
\xi_\ga(\eta) \sim \left(-\eta \right)^{\b \ga}, ~~~~~~~~~~~~~
-\eta_s<\eta <-\eta_1,
\label{44}
\eeq
during the high-curvature string phase described by the solution (\ref{26}). It reduces trivially to a constant in the final, post-big bang regime where the dilaton and the internal moduli are eventually stabilized, and where the canonical e.m. variable is freely oscillating with constant amplitude in Fourier space, $u_k'' +k^2 u_k=0$.

We are now in the position of determining the spectral distribution of the amplified e.m. fluctuations, and of computing in particular the spectral energy density of the photons produced from the vacuum, $\Om_\ga (\om, t_0)$, expressed in critical units and referred to the present fraction of radiation energy density, $\Om_r(t_0)$.

To this purpose we have to solve the canonical equation (\ref{31}) for the Fourier modes $u_k$ in the various phases of the pump field $\xi_\ga$, imposing the asymptotic normalization to an initial spectrum of quantum vacuum fluctuations, and matching the solutions at the transition epochs $\eta= -\eta_s$ and $\eta= -\eta_1$. Since there are only two relevant transitions, the final spectrum (for $\eta>-\eta_1$) will contain two branches: the high frequency modes with $\om_s<\om<\om_1$, leaving the horizon during the string phase and affected only by the final transition at $\eta=-\eta_1$, and the low frequency modes with $\om<\om_s$, leaving the horizon during the low energy phase and affected by both transitions. The contribution to $\Om_\ga$ of modes with $\om>\om_1$ turns out to be exponentially suppressed (see e.g. \cite{35}), and can be consistently neglected for the purpose of this paper. 

Following the procedure (and applying the results) of previous computations \cite{17,14,23} we then find: 
\bea
\Om_{\ga}(\om, t_0) &=& 
\Om_r (t_0) \left(H_1\over \Mp\right)^2 
\left(\om\over \om_1\right)^{3-|2\b\ga-1|},
~~~~~~~~~~~~~~~~~~~~~~~~~~ \om_s<\om<\om_1,
\nonumber \\ &=&
\Om_r (t_0) \left(H_1\over \Mp\right)^2 
\left(\om_s\over \om_1\right)^{3-|2\b\ga-1|}
\left(\om\over \om_s\right)^{3-\left|{\ga}{1-3 \b_0\over 1- \b_0}-1\right|},
~~~~~~~~ \om<\om_s. 
\label{45}
\eea
It is appropriate to recall, at this point, that the parameter $\b_0$ also controls the value of the scalar spectral index $n_s$, and that its value has to be fixed so as to 
satisfy Eq. (\ref{36}) (as discussed in section \ref{Sec3}). Hence, for a viable model of background, the spectral power of the low-frequency branch of the spectrum (\ref{45}) cannot be completely free, and can be expressed in terms of $n_s$ as follows:
\beq
\Om_r (t_0) \left(H_1\over \Mp\right)^2 
\left(\om_s\over \om_1\right)^{3-|2\b\ga-1|}
\left(\om\over \om_s\right)^{3-{1\over 2}\left|{\ga} (4-n_s)-2\right|},
~~~~~~~~~~~~ \om<\om_s. 
\label{46}
\eeq
It should be stressed, finally, that $\Om_{\ga}(\om, t_0)$ describes the spectral distribution of the total energy density of the photons produced by the inflationary amplification of the quantum fluctuations of the vacuum, including both electric and magnetic components, and evaluated for all modes present today inside the horizon.

Let us now impose the condition that the above spectrum of e.m. fluctuations contains magnetic modes $B_s$ of the appropriate amplitude and frequency scale   needed to seed the galactic dynamo, and to produce cosmic magnetic fields $B_G$ of current amplitude $B_G(t_0) \sim 10^{-6}$ Gauss. A conservative estimate of the required seeds then leads to a lower bound on the present amplitude of the magnetic fluctuations at the Mpc scale \cite{36}, which can be expressed as $B_s(\om_G, t_0) \gaq 10^{-23}$ Gauss, where $\om_G =1\, {\rm Mpc}^{-1}$. This imposes on the spectrum (\ref{45}), (\ref{46}) the constraint (first pointed out in \cite{36})
\beq
{B_s^2(\om_G, t_0) \over B_G^2(t_0)} \sim {\r_\ga (\om_G, t_0) \over \r_r(t_0)}
={\Om_\ga (\om_G, t_0) \over \Om_r(t_0)} \gaq 10^{-34},
~~~~~~~~~~~~~ \om_G =1\, {\rm Mpc}^{-1}
\label{47}
\eeq
(see also \cite{17,14,23}). To obtain this condition we have used the approximate equality of the energy density of the galactic magnetic field, $B_G^2$, and the total energy density of the cosmic radiation, $\r_r(t_0)$. Also, we have identified the spectral energy density of the e.m. fluctuations on the galatic scale, $\r_\ga (\om_G, t_0)$, with the contribution of its magnetic components, $B_s^2$.  Indeed, all frequency modes which today are freely oscillating inside the horizon, like the galactic scale $\om_G$, describe a freely propagating e.m. field, with the same content of electric and magnetic energy density (modulo the possible dissipation of the electric field when the Universe behaves like a highly conducting medium, see e.g. \cite{36}). We have thus, in particular, $\r_\ga (\om_G, t_0) \sim E_s^2(\om_G, t_0)  + B_s^2 (\om_G, t_0) \sim B_s^2(\om_G, t_0)$.

The above constraint on $\Om_\ga(\om_G)$ has to be complemented by a competing upper bound on the total energy density $\Om_\gamma(t)$ stored in the e.m. fluctuations, integrated over all modes and including both electric and magnetic components. Such a bound on $\Om_\gamma(t)$ must be satisfied both during and after inflation, and imposes on the seed energy density to be smaller than the background energy density to avoid destroying the large-scale homogeneity of the cosmic geometry, and to be consistent with the linearized treatment of the fluctuations as small perturbations with negligible back-reaction. This gives a condition which can be translated into a crude upper bound on the peak intensity of the spectrum, to be satisfied at all times, as follows \cite{12,17}:
\beq
\Om_\ga (\om_{\rm peak}, t_0) \laq \Om_r(t_0),
\label{48}
\eeq
where $\om_{\rm peak}$ is the corresponding peak frequency.

In order to satisfy the above limit, and to avoid an unbounded growth of the e.m. energy at low frequency (see also \cite{17a}), we shall restrict the analysis of this paper to a spectrum of e.m. fluctuations whose low-frequency branch ($\om <\om_s$) is always ``blue" (i.e., growing with frequency).  The high-frequency branch ($\om_s<\om<\om_1$), on the contrary, may grow or decrease with frequency: as a consequence, the peak of the spectrum may correspond either to $\om_1$ or to $\om_s$. 

For the explicit application of the constraints (\ref{47}), (\ref{48}) we must now specify the sign of the quantity $\ga(4-n_s)-2$, which controls the spectral power at low frequencies, according to Eq. (\ref{46}). We have in general two possibilities.

If $\ga \geq 2/(4-n_s)$ then the behavior of low-frequency (dilaton) branch of the spectrum is given by
\beq
\Om_\ga(\om) \sim \om^{4- \ga(4-n_s)/2}, 
~~~~~~~~~~~~~~ \om<\om_s,
\label{49}
\eeq
and the spectrum is blue for 
\beq
{2\over 4-n_s} \leq \ga <{8\over 4-n_s}.
\label{410}
\eeq
If, instead, $\ga \leq 2/(4-n_s)$ then the low-frequency spectral behavior becomes
\beq
\Om_\ga(\om) \sim \om^{2+\ga(4-n_s)/2}, 
~~~~~~~~~~~~~~ \om<\om_s,
\label{411}
\eeq
and the spectrum is blue for 
\beq
-{4\over 4-n_s}< \ga \leq {2\over 4-n_s}.
\label{412}
\eeq
The two conditions (\ref{410}), (\ref{412}) completely specify the allowed range of values for the parameter $\ga$ in the chosen model of background. We should recall that $\ga$ controls the time evolution of the pump field (\ref{42}), hence its value is crucial for the process of seeds production. 

As shown by an explicit computation, however, it turns out that for the case specified by Eq. (\ref{412}) an efficient production of seeds is only possible for negative valus of $\ga$: in other words, for an effective e.m. coupling $\xi_\ga^{-1}$ which decreases in time, according to Eq. (\ref{43}), (\ref{44}). This implies a very large value of the e.m. coupling at the beginning of the magnetogenesis process, which makes the treatment of the seeds as linear perturbations strongly inappropriate \cite{17a}. We shall thus omit the range of values (\ref{412}) from our subsequent discussion.

%%%%%%%%%%%%
 \subsection{Growing electromagnetic coupling}
\label{Sec41}
%%%%%%%%%%%%%%%%

We start considering the range of values of Eq. (\ref{410}), where $\ga$ is always positive and the  effective coupling associated to the e.m. fluctuations, $\xi^{-1}_\ga$, is  growing in time (see Eqs. (\ref{43}), (\ref{44}), (\ref{35})). In that case the frequency dependence of the dilaton branch of the spectrum is always given by Eq. (\ref{49}), but there are still  various possibilities to  consider, depending on the sign of the quantity  $2\b\ga-1$ (which controls the behavior of the string branch of the spectrum, see Eq. (\ref{45})), and depending on the relative localization of the frequency $\om_G$ in the two branches of the spectrum ($\om_G<\om_s$ or $\om_G>\om_s$). 

Let us first consider the case $2\b\ga \geq 1$, corresponding to a string branch with the spectral behavior $\Om_\ga \sim \om^{4-2\b\ga}$. If this spectrum is non-decreasing, i.e. if
\beq
1\leq 2 \b\ga \leq 4,
\label{413}
\eeq
then $\om_{\rm peak}=\om_1$, and the bound (\ref{48}) is always automatically satisfied. If the galactic scale belongs to the dilaton branch of the spectrum, $\om_G<\om_s$, namely if
\beq
z_s \left(H_1\over \Mp\right)^{-5/6} \laq 0.5 \times 10^{26}
\label{414}
\eeq
(see Appendix \ref{AppA}, Eq. (\ref{a6})), then the constraint (\ref{47}) gives 
\beq
z_s^{2\b\ga -\ga(4-n_s)/2} \left(H_1\over \Mp\right)^{2} 
\left[ 2 \times 10^{-26} \left(H_1\over \Mp\right)^{-5/6} \right]^{4-\ga(4-n_s)/2}
\gaq 10^{-34}.
\label{415}
\eeq
If, on the contrary, $\om_G>\om_s$, then Eq. (\ref{47}) implies
\beq
 \left(H_1\over \Mp\right)^{2} 
\left[ 2 \times 10^{-26} \left(H_1\over \Mp\right)^{-5/6} \right]^{4-2 \b\ga}
\gaq 10^{-34}.
\label{416}
\eeq
Finally, if the string branch has a red spectrum, namely if
\beq
2\b\ga>4,
\label{417}
\eeq
then the peak is localized at $\om_{\rm peak}=\om_s$, the above constraints (\ref{414}) and (\ref{415}) are still valid, but they are to be supplemented by the upper limit (\ref{48}), which implies
\beq
z_s^{2\b\ga-4}  \left(H_1\over \Mp\right)^{2} <1.
\label{418}
\eeq

The above analysis has to be repeated for the (alternative) possibility that
\beq
2 \b\ga <1,
\label{419}
\eeq
corresponding to a string branch with the spectral behavior $\Om_\ga \sim \om^{2+2\b \ga}$ (see Eq. (\ref{45})). This spectrum is always blue (as $\b>0$, $\ga >0$), the peak  is always at $\om_1$, and the bound (\ref{48}) is automatically satisfied. In the case of Eq. (\ref{414}), i.e. $\om_G<\om_s$, the constraint (\ref{47}) gives
\beq
z_s^{2-2\b\ga -\ga(4-n_s)/2} \left(H_1\over \Mp\right)^{2} 
\left[ 2 \times 10^{-26} \left(H_1\over \Mp\right)^{-5/6} \right]^{4-\ga(4-n_s)/2}
\gaq 10^{-34},
\label{420}
\eeq
while in the opposite case $\om_G>\om_s$ the constraint (\ref{47}) gives
\beq
 \left(H_1\over \Mp\right)^{2} 
\left[ 2 \times 10^{-26} \left(H_1\over \Mp\right)^{-5/6} \right]^{2+2 \b\ga}
\gaq 10^{-34}.
\label{421}
\eeq

\begin{figure}[t]
\centering
\includegraphics[width=9cm]{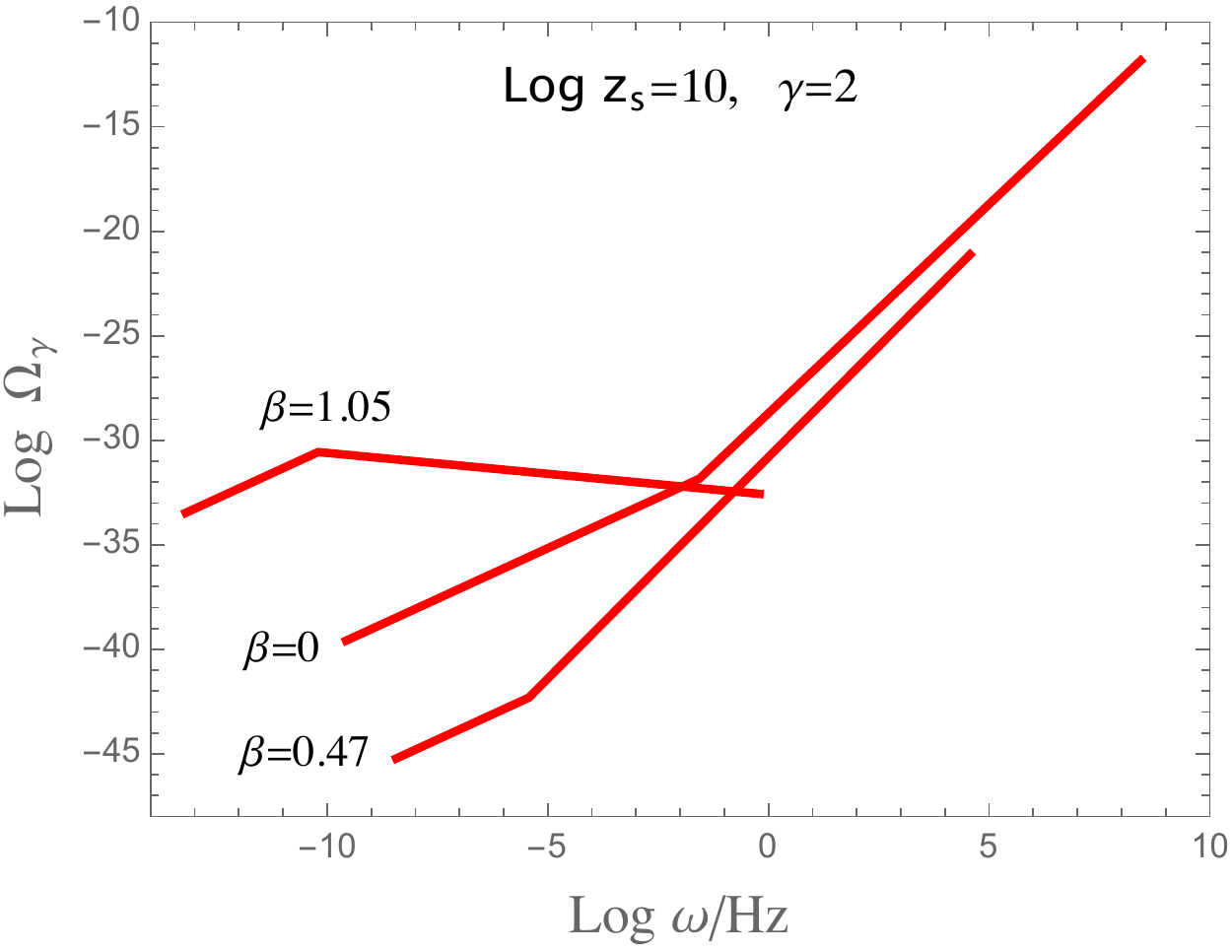}
\caption{A typical example of the seed spectrum (\ref{45}), (\ref{46}), with $\ga=2$ and with $H_1$ satisfying the CMB constraint (\ref{37}). We have used $n_s=0.968$, $\log \Om_r=-4$, and we have plotted the spectrum for 
three possible values of the string parameter: 
$\b=0, \b=0.47, \b=1.05$. In the low-frequency (dilaton) branch the spectral slope is the same for all $\b$ (and only controlled by $\ga$, according to Eq. (\ref{46})). In the high-frequency (string) branch we have instead a strong $\b$-dependence, and also the possibility of a red spectrum, according to Eq. (\ref{417}).} 
\label{f2}
\end{figure}

Let us now combine all cases with their associated constraints, and ask whether there are values of $\ga$ included in the allowed interval (\ref{410}) which may consistently satisfy such constraints, at least in some of the cases listed before. Let us impose, also, that the production of seeds is compatible with the bounds imposed by CMB observations, and in particular with the conditions given by Eqs. (\ref{27}), (\ref{211}), (\ref{212}), (\ref{37}) (determining the light-blue allowed region of Fig. \ref{f1}). 

In such a case we may rewrite all constraints (\ref{413})--(\ref{421}) by eliminating $H_1/\Mp$ through Eq. (\ref{37}), and by expressing $z_s$ and $\b$ in terms of $x$ and $y$ according to eq. (\ref{314}). By plotting the resulting allowed region in the plane $\{x,y\}$ we find a non-vanishing area only for values of $\ga$ lying in a restricted portion of the interval (\ref{410}), namely for
\beq
1.4 \laq \ga <{8\over 4-n_s}.
\label{422}
\eeq
Notice that the case $\ga=1$, i.e. the typical value of $\ga$ predicted by the heterotic string model at the tree-level, is outside this allowed range.

\begin{figure}[t]
\centering
\includegraphics[width=9cm]{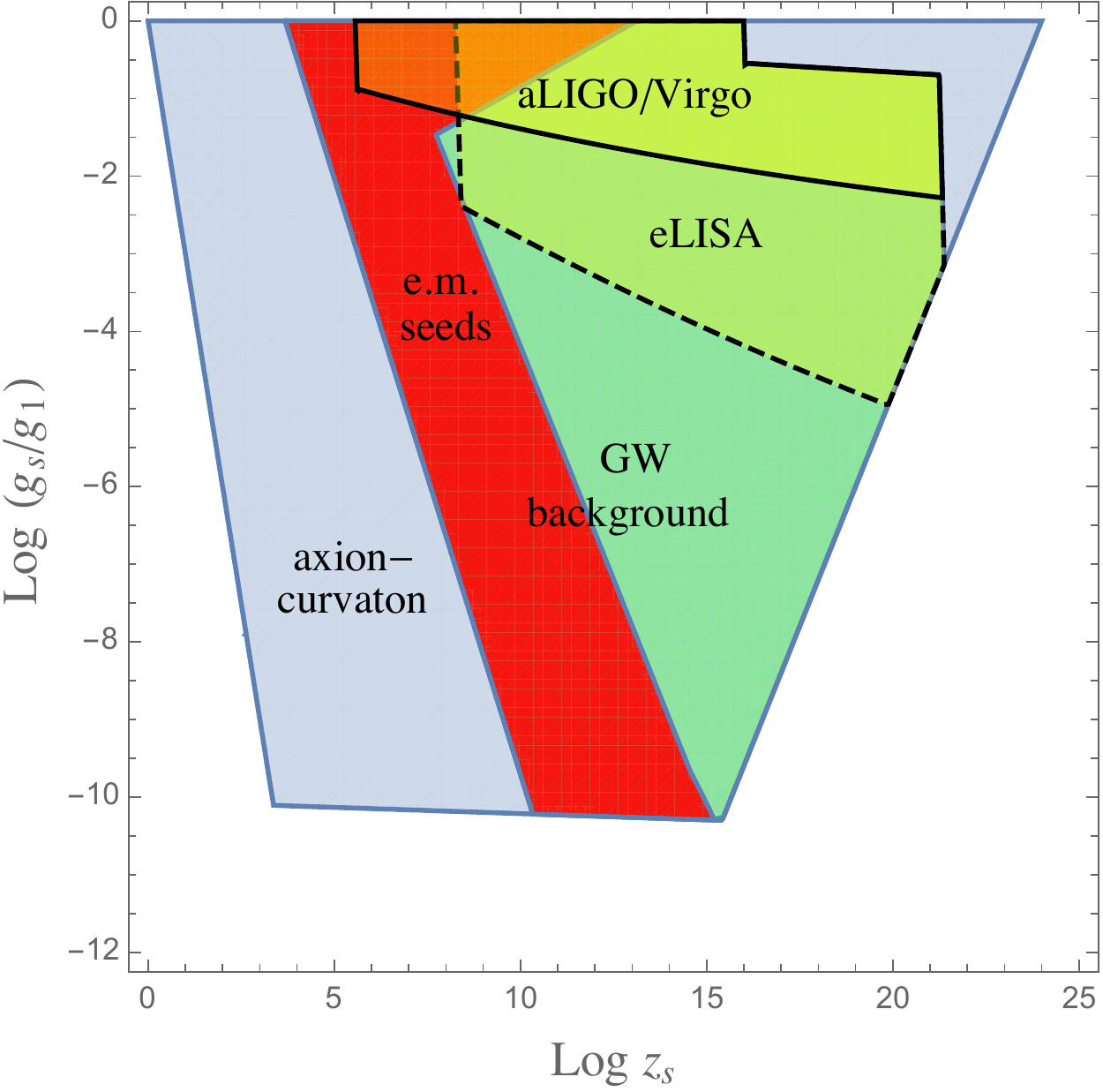}
\caption{Same as Fig. \ref{f1}. In addition we have illustrated, with the red area, the  region of parameter space where an efficient production of magnetic seeds is compatible with the bounds imposed by scalar and tensor perturbations. The allowed values of $\ga$, for that region, are varying in the range of Eq. (\ref{423}). Such a region has a non-vanishing intersection (represented by the orange area) with the portion of parameter space where the stochastic GW background is detectable by the aLIGO/Virgo network and by eLISA.} 
\label{f3}
\end{figure}

If we impose, in addition, that all the constraints associated with the production of a viable GW background (i.e. those determining the green area of Fig. \ref{f1}) are also satisfied, then we find that the allowed values of $\ga$ are further reduced, and limited to the range 
\beq
2 \laq \ga <{8\over 4-n_s}.
\label{423}
\eeq
An example of the resulting seed spectrum, for the typical case $\ga=2$, is illustrated in Fig. \ref{f2}. 
For the values of $\ga$ satisfying Eq. (\ref{423}) we have a class of models in which the inflationary amplification of the e.m. fluctuations can efficiently produce seeds for the galactic magnetic fields, and satisfy all phenomenological bounds following from the associated amplification of scalar and tensor perturbations. The corresponding allowed region of parameter space is illustrated by the red area plotted in Fig. \ref{f3}. It should be noted that such a region is also compatible -- in a restricted range of parameters, represented by the orange area -- with the production of a stochastic GW background possibly detectable by the interferometers aLIGO/Virgo and eLISA.

%%%%%%%%%%%%%%%%%%%%%%%%%%%%%%%%%%%%%%%%%%%%%%%
\section{Conclusion}
\label{Sec5}
\setcounter{equation}{0}
%%%%%%%%%%%%%%%%%%%%%%%%%%%%%%%%%%%%%%%%%%%%%%%

In this paper we have considered a class of string cosmology models whose spectrum of scalar and tensor perturbations can reproduce the current observational data and satisfy all known phenomenological constraints. In that context, we have studied the amplification of the e.m. fluctuations using a one-parameter model of e.m. pump field, possibly describing different string-theory representations of the e.m. interactions and/or different schemes of dimensional reduction. 

We have found, in such a context, that the possibility of producing primordial seeds for the cosmic magnetic fields is severely affected by the constraints imposed by scalar and tensor perturbations. In particular, the allowed region for an efficient seed production, in parameter space, in non-vanishing only for a small, limited range of values of the parameter $\ga$ (see Eqs. (\ref{423})), controlling the coupling of photons to the background geometry.

The interesting result is that such an allowed region has a significant intersection with the region of parameter space where the primordial GW background is strong enough to be detectable by near-future experiments, planned by the interferometers aLIGO/Virgo and eLISA (see Fig. \ref{f3}). This implies that forthcoming data from those interferometers can also provide information -- even if in a model-dependent way -- on the inflationary mechanism of seeds production. In particular, a positive signal from the GW detectors, able to select a localized portion of parameter space, could directly confirm (or exclude) the possibility that efficient magnetic seeds have been amplified by a phase of pre-big bang inflation, as described by the class of models we have considered.

%%%%%%%%%%%%%%%%%%%%%%%%%
% \vskip 1 cm
 
\section*{Acknowledgements}

I wish to thank Gabriele Veneziano for many useful discussions that have motivated, in part, the preparation of this paper. This work is supported in part by MIUR under grant no. 2012CPPYP7 (PRIN 2012), and by INFN under the program TAsP (Theoretical Astroparticle Physics).

%%%%%%%%%%%%%%%%%%%%%%%%%%%%%%%
%%%%%%%%%%%%%%%%%%%%%%%%%%%%%%%
\appendix
%%%%%%%%%%%%%%%%%%%%%%%%%%%%%%%%%%%

\section{A relevant frequency ratio}
\label{AppA}
\setcounter{equation}{0}
%%%%%%%%%%%%%%%%%%%%%%%%%%%%%%%%%%%%%%%%

We shall present here a detailed computation of the frequency ratio $\om_G/\om_s$, whose explicit value is relevant for the discussion of seeds production in the model of background of this paper. The same procedure can also be applied to compute other frequency ratios, relevant to our discussion and to our model of background evolution. 

Let us start with the definition of the string redshift parameter, $z_s= a_1/a_s=\om_1/\om_s$ (see section \ref{Sec2} and section \ref{Sec3}), which implies
\beq
{\om_G\over \om_s}= {\om_G\over \om_1}{\om_1\over \om_s}= z_s \,{\om_G\over \om_1}.
\label{a1}
\eeq
Also, let us recall that $\om_1$ is the proper frequency of a mode crossing the horizon at the end of the string phase, i.e. at the transition scale $H_1$ such that $\om_1(t)= H_1 a_1/a(t)$, while $\om_G$ is the proper frequency of a mode re-entering the horizon at the curvature scale $H_G$, such that $\om_G(t)= H_{G} a_G/a(t)$. Hence:
\beq
{\om_G\over \om_s}=  z_s {H_G a_G\over H_1a_1}.
\label{a2}
\eeq

It should be noted, at this point, that the scale of interest $\om_G= 1\,{\rm Mpc}^{-1}$ re-enters the horizon when the Universe is still radiation-dominated (since $H_G$ is larger than the equality scale, $H_G>H_{\rm eq}$). Since our model of post-big bang evolution is initially of the matter-dominated type ($a \sim \eta^2\sim H^{-2/3}$) from $H_1$ down to the axion decay scale $H_d$, and of the radiation-dominated type ($a \sim \eta\sim H^{-1/2}$) from $H_d$ down to $H_G$ (see section \ref{Sec2}), we can thus write:
\beq
{\om_G\over \om_s}=  z_s {H_G \over H_1}\left(a_G\over a_d\right)_{\rm rad} 
\left(a_d\over a_1\right)_{\rm mat} =  z_s {H_G \over H_1}
\left(H_d\over H_G\right)^{1/2}
\left(H_1\over H_d\right)^{2/3}.
\label{a3}
\eeq
Using Planck units and recalling that, for our background, $H_d/\Mp= (H_1/\Mp)^3$, we obtain:
\beq
{\om_G\over \om_s}=
 z_s 
\left(H_G\over \Mp\right)^{1/2}
\left(H_1\over \Mp\right)^{-5/6}.
\label{a4}
\eeq
Finally, we may conveniently refer the galactic scale to the matter-radiation equality scale by noting that $\om_G =1\,{\rm Mpc}^{-1} \sim 10^2 \,\om_{\rm eq}$, where $\om_{\rm eq}$ is the proper frequency of a mode re-entering the horizon at the equality epoch. This gives
\beq
{\om_G\over \om_{\rm eq}}=\left(H_G\over H_{\rm eq}\right)^{1/2} \sim 10^2,
\label{a5}
\eeq
so that, using $H_{\rm eq} / \Mp \simeq 0.6 \times 10^{-55}$, we eventually obtain
\beq
{\om_G\over \om_s}=2\times 10^{-26}\,  z_s 
\left(H_1\over \Mp\right)^{-5/6}.
\label{a6}
\eeq

%%%%%%%%%%%%%%%%%%%%%%%%%%%%
%%%%%%%%%%%%%%%%%%%%%%%%%%%%%%

\end{document}